\documentclass[twocolumn,showpacs,preprintnumbers,amsmath,amssymb]{revtex4}

\usepackage{graphicx}
\usepackage{dcolumn}
\usepackage{bm}

\begin{document}


\title{Experimental demonstration of a high speed quantum random number generation scheme based on measuring phase noise of a single mode laser}

\author{Bing Qi, Yue-Meng Chi, Hoi-Kwong Lo, Li Qian}
\affiliation{Center for Quantum Information and Quantum Control,
Dept. of Electrical and Computer Engineering and Dept. of Physics,
University of Toronto, Toronto, Ontario, M5S 3G4, Canada
}%

\date{\today}

\begin{abstract}
We present a high speed random number generation scheme based on
measuring the quantum phase noise of a single mode diode laser
operating at a low intensity level near the lasing threshold. A
delayed self-heterodyning system has been developed to measure the
random phase fluctuation. By actively stabilizing the phase of the
fiber interferometer, a random number generation rate of 500Mbit/s
has been demonstrated and the generated random numbers have passed
all the DIEHARD tests.
\end{abstract}

\maketitle

\section{Introduction}

Random numbers have been widely used in many branches of science and
technology, such as statistical analysis, computer simulation
\cite{Monte_Carlo}, cryptography \cite{Schneier}, etc. One recent
example is quantum key distribution (QKD) \cite{BB84}, where truly
random numbers are required for both quantum state preparation and
quantum state detection. Most recently, truly random numbers have
also been employed in testing fundamental principles of physics
\cite{Jacques07, bell}.

In practice, it is not easy to obtain high quality random numbers
with proven randomness \cite{Barak03}. In a cryptographic system,
the application of a weak random number generator (RND) could be
catastrophic, as evidenced by Goldberg and Wagner's attack on the
Netscape SSL implementation \cite{GW96}.

A pseudorandom generator generates a long train of ``random'' bits
from a short random seed by employing deterministic algorithms. The
generated long bit string could meet a number of statistical
measures, which allows it to pass all existing randomness tests.
However, the entropy of the long bit string is ultimately determined
by the length of the random seed. In principle, random numbers
generated by deterministic algorithms are not truly random. John von
Neumann once famously said ``Anyone who considers arithmetical
methods of producing random digits is, of course, in a state of
sin.''\cite{Neumann51}.

A physical RNG, on the other hand, generates random numbers from
unpredictable physical processes, such as thermal noise
\cite{thermal}, radioactive decay \cite{Hotbit}, air turbulence
\cite{Davis94}, etc. For example, the Intel 80802 Firmware Hub chip
included a hardware random number generator \cite{Intel}. It is
important to distinguish two different types of physical random
number generation processes, based on the source of randomness: the
chaotic behavior of classical deterministic systems, which we shall
call type-one randomness henceforth; or the truly probabilistic
nature of fundamental quantum processes \cite{Jennewein00}, which we
call type-two randomness. In this paper, we use the term ``classical
noise'' to represent the unpredictability of a deterministic chaotic
system and the term ``quantum noise'' to describe the fundamental
uncertainty in a quantum process.

For example, a RNG based on atmospheric conditions can be treated as
a type-one RNG, since the randomness mainly originates from the
absence of enough information about the weather system. In other
words, the observed fluctuation can be treated as classical noise.
Obviously, the more knowledge we have, the less random the system
appears. The unpredictable weather change appeared to a layman might
well be predictable to an expert equipped with supercomputers. This
raises a serious question: how much can we trust a RNG? The weakness
of a RNG could be fatal if the generated random numbers have been
used as secure keys in a cryptographic system where the security
relies on the true randomness of the key.

We remark that ultrahigh speed RNGs based on chaotic semiconductor
lasers have been proposed and random number generation rates above
Gbit/s have been demonstrated \cite{Uchida08}. In \cite{Uchida08},
random numbers are generated from the amplitude noise of chaotic
semiconductor lasers. However, as we discussed above, it is still
arguable whether random numbers generated from a deterministic
chaotic system are suitable for cryptographic applications.

Fortunately, a type-two RNG, or a quantum RNG (QRNG), can provide us
with true random numbers with proven randomness. One of the simplest
QRNG is constructed by a single photon source, a symmetric beam
splitter and two single photon detectors \cite{Jennewein00}. Each
photon has the same probability to be either transmitted or
reflected by the beam splitter and thus ``which detector clicks'' is
completely unpredictable. This unpredictability is not due to the
absence of information about the quantum state of the photon or the
measurement device, it is due to the probabilistic nature of a
projection measurement. The randomness of the result is guaranteed
by the fundamental laws of quantum mechanics.

One natural source of quantum randomness is the thermal noise in an
electrical amplifier \cite{thermal}. However, a practical high gain,
broadband electrical amplifier also exhibits classical noises which
typically dominate over the quantum noise. So far, the reported
random number generation rate based on this scheme is only a few
Mbit/s \cite{thermal}.

To date, most QRNGs are based on performing single photon detections
\cite{Jennewein00,Dynes08} and the highest random number generation
rate achieved is 16 Mbit/s \cite{idq_RNG}. This random number
generation rate is too low for certain applications, such as high
speed QKD systems operated at GHz clock rates
\cite{Takesue07,Yuan08}. Indeed, current high-speed QKD set-ups
often use either a) deterministic random number generation
algorithms or b) repeatedly a fixed pattern and, therefore, are not
really unconditionally secure. Although there is still some room for
improvement, the ultimate speed of these devices is limited by the
performance of the single photon detector (SPD), such as its dead
time, efficiency, afterpulsing probability, etc. For example, a
typical silicon APD-SPD has a dead time of tens of ns \cite{SPCM},
which suggests that the ultimate random number generation rate based
on this type of SPD is in the order of tens of Mbit/s.

Another promising quantum random number generation scheme is based
on measuring the random field fluctuation of vacuum with a homodyne
detector \cite{Trifonov07}. However, the fabrication of a high
speed, shot noise limited homodyne detector is also a challenging
task and a high speed QRNG ($>$ tens of Mbit/s) based on this scheme
has not been demonstrated.

In this paper, we present a QRNG scheme based on measuring the
quantum phase noise of a single mode semiconductor laser operating
at a low intensity level near the lasing threshold \cite{AQIS09}.
This ensures that the main contribution to the phase noise is from
spontaneous emission (SE)\cite{Henry82}, rather than from chaotic
evolution of the macroscopic field \cite{Uchida08}. One significant
advantage of this scheme is the potential high random number
generation rate. In this paper, we achieve a 500Mbit/s random number
generation rate with commercial off-the-shelf components.

We remark that any practical devices present both quantum noise and
classical noise, and true random numbers can only originate from the
former one. How can we distinguish these two different noises in
practice? We will address this question briefly in Section V.

\section{Theoretical model}

\subsection{Quantum phase noise and linewidth of semiconductor
lasers}

The linewidth of a single mode semiconductor laser can be viewed as
due to the random phase fluctuation of the optical field
\cite{Henry82}. Experimental studies have shown that the linewidth
of a single mode injection diode laser varies linearly with
reciprocal laser output power \cite{Fleming81,Yamamoto81}. To
explain these experimental discoveries, Henry developed a
theoretical model which attributes the fundamental laser linewidth
to the phase fluctuations arise from spontaneous emission
\cite{Henry82}. Furthermore, it has been shown that the phase noise
of an InGaAsP DFB laser can be well described by this model
\cite{Tkach86}.

In \cite{Henry82}, the dependence of the linewidth of a single mode
semiconductor laser on its output laser power is described by
\begin{equation}
\Delta f=\frac{\nu_{g}^{2}h\nu
g\eta_{sp}\alpha_{m}(1+\alpha^{2})}{8\pi P_{0}}
\end{equation}
Here, $\nu_{g}$ is the group velocity, $h\nu$ is the energy of
photon, $g$ is the gain of laser medium, $\eta_{sp}$ is the
spontaneous emission factor, $P_{0}$ is the output power per facet.
$\alpha_{m}$ is the facet loss which is defined as $\alpha_{m}\equiv
g-\alpha_{L}$, where $\alpha_{L}$ is the waveguide loss of the
laser. $\alpha\equiv\frac{\Delta n'}{\Delta n''}$ , where $\Delta
n'$ and $\Delta n''$ are the deviations of the real part and
imaginary part of the refractive index from their steady-state
values.

An intuitive physical picture is as follows \cite{Henry82}: each
spontaneous emitted photon has a random phase, which in turn
contributes a random phase fluctuation to the total electric field
and results in a linewidth broadening. This is represented by the
term ``1'' in the parentheses on the right hand side of (1). On the
other hand, the same spontaneous emitted photon also alters the
amplitude of the laser field, which results in a change of the
carrier density. The change of carrier density further triggers a
change of $n''$, which is the imaginary part of the refractive index
of the laser medium. Finally, the change in $n''$ has an associated
change of the real part of the refractive index $n'$, which
contributes to an additional phase shift of the laser field and
linewidth broadening. This additional linewidth broadening is
described by the term $\alpha^2$ in the parentheses on the right
hand side of (1).

Note that the spontaneous emission is a quantum mechanical effect
and the corresponding phase noise can be treated as quantum noise.
However, a practical laser source also presents additional classical
noises, such as occupation fluctuation \cite{Vahala83} and 1/f noise
\cite{Kikuchi83}. Fortunately, these classical noises are laser
power independent \cite{Vahala83,Kikuchi83}. By operating a
semiconductor laser under certain power level, the noise properties
are mainly determined by quantum effects \cite{Vahala83}. In the
following Sections we will show how to harness this quantum noise to
generate true random numbers.

\subsection{Phase measurement with a delayed self-heterodyning scheme}

The optical phase of a laser field can be measured by performing an
interferometric experiment. A delayed self-heterodyning scheme has
been employed to measure the linewidth of a semiconductor laser
\cite{Yamamoto81}. Fig.1 shows its basic structure.

\begin{figure}[!t]\center
\resizebox{8.5cm}{!}{\includegraphics{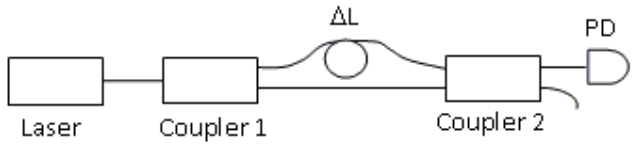}}
\caption{The basic structure of a fiber based delayed
self-heterodyning system. $\Delta L$ is the path length imbalance;
PD is a photo-detector.}
\end{figure}

The electric field of a laser beam can be described by
\begin{equation}
E(t)=E_0exp[i(\omega_0 t+\theta(t))]
\end{equation}
where $\theta(t)$ represents the random phase fluctuation of the
laser source.

The interference signal detected by the photo detector (PD) can be
described by
\begin{equation}
S(t)\propto |E_0exp[i(\omega_0 t+\theta(t+T_d))]+E_0exp[i(\omega_0
t+\omega_0T_d+\theta(t))]|^2
\end{equation}
Here $T_d$ is the time delay difference between the two arms of the
Mach-Zehnder interferometer (MZI), which can be determined by
$T_d=n\Delta L/C$. $\Delta L$ is the path length imbalance, $n$ is
the refractive index of fiber and $C$ is the speed of light in
vacuum.

After removing a DC background, equation (3) can be simplified as
\begin{equation}
S(t)\propto cos[\omega_0 T_d +\Delta\theta(t,T_d)]
\end{equation}
where $\Delta\theta(t,T_d)\equiv\theta(t)-\theta(t+T_d)$.

In (4), the term $\Delta\theta(t,T_d)$ represents the quantum phase
noise of the laser source, while the term $\omega_0 T_d$ represents
the phase delay introduced by the path length difference. In
practice, for an interferometer without phase stabilization, the
term $\omega_0 T_d$ will not be a constant due to ambient
temperature fluctuations, for example. This in turn will contribute
additional classical phase noise. Intuitively, if the time delay
difference $T_d$ is much larger than the coherence time of the
laser, then $\Delta\theta(t,T_d)$ will present an uniform
distribution in the range of $[-\pi,\pi)$. Under this condition, the
total phase $\omega_0 T_d +\Delta\theta(t,T_d)$ also uniformly
distributes in the range of $[-\pi,\pi)$ regardless of the actual
value of $\omega_0 T_d$. Thus we can generate binary random numbers
by simply measuring $S(t)$ using a fast detector, sampling the
output at the fixed intervals ($T_S$) to generate a series of
$\widetilde{S}(t_i)$, and taking the sign of the individual
$\widetilde{S}(t_i)$ in the series.

A more rigorous discussion is as follows: the net contribution of a
large number of SE photons can be characterized by a random walk
process, and the phase fluctuation $\Delta\theta(t,T_d)$ can be
treated as Gaussian white noise with a variance of \cite{Yariv07}:
\begin{equation}
\langle[\Delta\theta(t,T_d)]^2\rangle=\frac{2T_d}{\tau_c}
\end{equation}
Here $\tau_c$ is the coherence time of the laser, which is related
to its linewidth $\Delta f$ as $\tau_c\simeq\frac{1}{\pi\Delta f}$
\cite{Yariv07}.

Equation(5) shows that as long as $T_d\gg\tau_c$, the resulting
Gaussian distribution can be treated as a uniform distribution in
the range of $[-\pi,\pi)$ in practice.

It is useful to define two other time constants here. The response
time of the photo-detection system $T_R$ is defined as the
reciprocal of the detection system's bandwidth. The sampling period
$T_S$ is defined as the reciprocal of the sampling rate.

The necessary condition for random number generation without phase
stabilization is summarized as
\begin{equation}
T_d\gg\tau_c; T_S-T_d\gg\tau_c+T_R
\end{equation}

From (6), the maximum sampling rate (or the random number generation
rate) is determined by the coherence time of the laser. The
coherence time of the laser used in our experiment can be set to a
few ns by tuning its driving current. To generate high quality
random numbers, the sample period $T_S$ should be larger than 10ns,
which corresponds to a maximum sample rate of 100MHz.

One way to go beyond the limitation imposed by the coherence time of
the laser source is to employ phase stabilization technique. This
can be seen from (4). By stabilizing the phase of the MZI, the term
$\omega_0 T_d$ in the cosine function can be treated as a constant.
Furthermore, if we can set $\omega_0 T_d=2m\pi+\pi/2$ (where $m$ is
an integral), equation (4) can be further simplified as
\begin{equation}
S(t)\propto sin[\Delta\theta(t,T_d)]
\end{equation}
Note in the derivation of (7), we have ignored the error of the
phase feedback control system, which will contribute to additional
classical noise. It will be interesting to quantify how the
performance of the proposed RNG depends on the phase control error.

From equation (7), the discrete time series samples of $S(t)$, which
is labeled as $\widetilde{S}(t_i)$, has a symmetric distribution
around zero. Again, we can generate binary random numbers by simply
taking the sign of $\widetilde{S}(t_i)$. Here, we don't need to
assume that $\Delta\theta(t,T_d)$ is uniformly distributed in the
range of $[-\pi,\pi)$. In principle, the sampling rate is mainly
limited by the bandwidth of the detection system but not the
coherence time (or linewidth) of the laser.

To minimize the correlation between adjacent samples, the time delay
imbalance $T_d$ should be smaller than the sampling period $T_S$.
This is illustrated in Fig.2: suppose that the first sampling result
$\widetilde{S}(t_1)$ is determined by phase noise from the SE
photons emitted in the time period of ($t_1-T_d$, $t_1$), while the
second sampling result $\widetilde{S}(t_2)$ is determined by phase
noise from the SE photons emitted in the time period of ($t_2-T_d$,
$t_2$). By choosing $T_S=t_2-t_1>T_d$, $\widetilde{S}(t_1)$ and
$\widetilde{S}(t_2)$ are contributed by SE photons emitted at
different time windows, thus there is no correlation between them.

\begin{figure}[!t]\center
\resizebox{6.5cm}{!}{\includegraphics{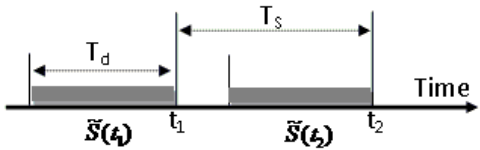}}
\caption{The relation between the time delay imbalance $T_d$ and the
sampling period $T_S$.}
\end{figure}

In practice there are other factors to be considered on determining
the optimal coherence time. On one hand, the coherence time $\tau_c$
should be larger than the response time $T_R$ of the detection
system. Otherwise, the interference signal will be averaged out due
to its random fluctuation within the time period for acquiring one
sample. On the other hand, from (5), $\tau_c$ cannot be too large,
otherwise, the variance of phase noise $\Delta\theta(t,T_d)$ could
be too small to be resolved.

The necessary condition for random number generation with phase
stabilization is summarized as
\begin{equation}
T_S-T_d>T_R
\end{equation}

\section{Experimental Setup}

A 1.5 $\mu m$ single mode cw DFB diode laser (ILX Lightwave) is
employed as the source of quantum phase noise. From (1), the
linewidth of the laser diode varies linearly with reciprocal laser
output power which can be conveniently controlled by adjusting the
driving current. By operating the laser diode at a power level where
its linewidth is much larger than the linewidth at the high power
limit, the majority of the phase noise can be attributed to quantum
noise.

A delayed self-heterodyning system has been developed to measure the
random phase fluctuation of the single mode DFB diode laser.  The
experimental setup is shown in Fig.3.  Two symmetric fiber couplers
are used to construct a fiber MZI with a length imbalance of $\Delta
L$. The interference signals from the two output ports of the second
fiber coupler are fed into two detection channels: the first
channel, including a 5GHz bandwidth InGaAs photo-detector ($PD_1$ in
Fig.3) and a 3GHz bandwidth real time oscilloscope, is used to
generate random numbers; the second channel ($PD_2$ in Fig.3), which
has a bandwidth of 1MHZ, is used to monitor the relatively slow
phase drift of the MZI due to ambient fluctuations. Due to its small
bandwidth, $PD_2$ can only sense the slow phase drift of MZI, while
the high frequency phase fluctuation due to SE will be averaged out.
The output from $PD_2$ is sampled by a computer together with a DAQ
card (NI PCI6115), which in turn provides a feedback control signal
to a phase modulator inside the MZI. Two polarization controllers
are used in this setup: $PC_1$ is used to make sure the polarization
state of the input light aligned with the axis of the phase
modulator, while $PC_2$ is employed to achieve high interference
visibility.

\begin{figure}[!t]\center
\resizebox{8.5cm}{!}{\includegraphics{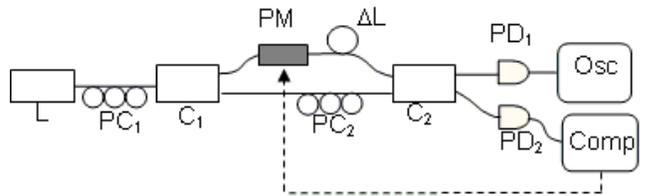}}
\caption{Experimental setup. $L$-1550nm DFB diode laser;
$PC_1,PC_2$-polarization controllers; $PM$-phase modulator;
$C_1,C_2$-fiber couplers; $PD_1$-5GHz photo-detector for random
number generation; $PD_2$-1MHz photo receiver for phase monitoring;
$Osc$-3GHz real time oscilloscope; $Comp$-Desktop computer with a NI
PCI6115 DAQ card; the length imbalance of the MZI is $650\pm100$
ps.}
\end{figure}

\section{Experimental results}

As we have discussed in Section II, by stabilizing the phase of MZI
to satisfy the condition of $\omega_0 T_d=2m\pi+\pi/2$ (where $m$ is
an integral), the sampling rate is not limited by the coherence time
of the laser, so we can achieve a very high random number generation
rate.

During this experiment, the driving current of the DFB laser has
been set to $I=12mA$. Note, if the driving current is too small, the
laser power will be too weak and the detected signal will be
dominated by the noise of the detection system rather than the
quantum phase noise of the laser; on the other hand, if the driving
current is too large, the quantum phase fluctuation could be too
small to be resolved. By using the technique described in
\cite{Yamamoto81}, the coherence time $\tau_c$ of the laser has been
determined to be about 10ns (or a linewidth of 30MHz) under the
condition of $I=12mA$ and 320ns (or a linewidth of 1MHz) at a high
driving current ($I=50mA$). From (5), for a fixed $T_d$, the phase
noise variance is proportional to $1/\tau_c$. Thus the phase noise
variance at $I=12mA$ is about 32 times larger than that at $I=50mA$.
Since we attribute the phase noise variance at the high power limit
(or the high driving current) to classical noise and assume that it
is laser power independent, we conclude that under our experimental
conditions, the phase noise of the laser is dominated by quantum
noise.

For an ideal RNG, there is no correlation between its outputs at
different times.  From the correlation theorem in Fourier
transformation, the spectrum of an ideal RNG is expected to be flat
\cite{FFT}.

We have measured the noise spectrum of the setup shown in Fig.3. In
this experiment, a spectrum analyzer (HP8564E) has been employed to
replace the oscilloscope. Measurements have been performed with two
different imbalanced MZIs, $T_{d1}=650\pm100ps$ and
$T_{d2}=250\pm100ps$. The measurement results are shown in Fig.4.
The electrical noise of the detection system has been measured by
blocking the laser output.

\begin{figure}[!t]\center
\resizebox{9.5cm}{!}{\includegraphics{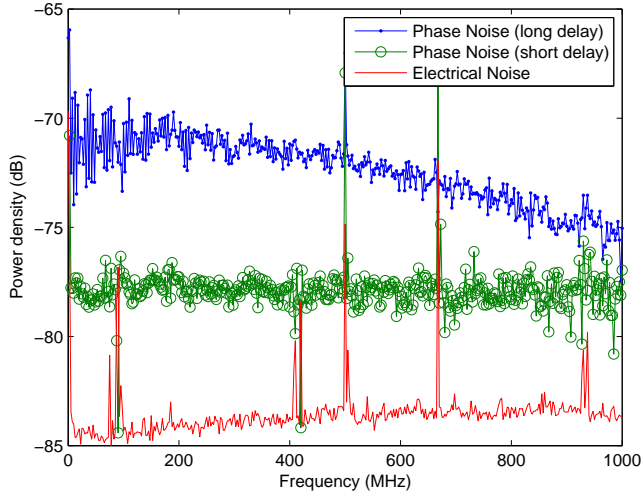}}
\caption{Spectral power density of electrical and phase noise. The
solid-line, dot-line and circle-line represent the spectral power
densities of the detection system, the phase noise with a long delay
($T_d=650\pm100ps$), and the phase noise with a short delay
($T_d=250\pm100ps$), correspondingly.}
\end{figure}

There are several remarkable features in Fig.4. First of all, the
electrical noise, which looks quite random in time domain, presents
a few dominant spectral lines. These spikes could be due to the
environmental EM noises picked up by our detection system. This
highlights the challenge in random number generation from the
thermal noise of a broadband electrical amplifier: the residual
classical noises could dominate over the quantum noise. On the other
hand, the spectra of phase noise are broadband as expected.
Secondly, the noise spectrum measured with a $T_d$ of $650\pm100ps$
presents a clear low-pass character, while the noise spectrum
measured with a smaller $T_d$ ($250\pm100ps$) shows a flatter
frequency response. As we have shown in Fig.2, the measured phase
noise is contributed by the SE photons emitted in a time interval of
$T_d$. This introduces an equivalent integration time in the order
of $T_d$ and thus reduces the bandwidth of the whole system to
$~1/T_d$. Third, at low frequency region, the noise power measured
at $T_d=250\pm100ps$ is about 7dB lower than that at
$T_d=650\pm100ps$. This is consistent with (5), which suggests a
linear relation between phase variance and the time delay $T_d$.

To generate random numbers, $T_d$ was set to be $650\pm100ps$ and
the output of PD1 was sampled by the 3GHz oscilloscope at a sampling
rate of 1G samples per second (corresponding to $T_S=1ns$). The
sampling results were saved onto the hard drive of the oscilloscope
in frames: the oscilloscope continuously samples 1M data, transfers
the data into the hard drive and then starts to sample another frame
of data. To generate binary random numbers from the sampling results
$\widetilde{S}(t_i)$, we simply compare them with the mean value
$S_0$: the $i_{th}$ bit is assigned as either ``1'' if
$\widetilde{S}(t_i)>S_0$ or ``0'' if $\widetilde{S}(t_i)<S_0$.

Two 100Mbits binary random number trains have been generated, which
are named as $Bin1$ and $Bin2$. By performing a bitwise XOR
operation between $Bin1$ and $Bin2$, the randomness can be further
improved. This XOR operation has been commonly used on improving
randomness of a RNG \cite{Epstein03,Barak03}. The random number
train generated through this XOR process is named as $Bin3$. Since
$Bin1$ and $Bin2$ have been generated at 1Gbit/s, the equivalent
generation rate of $Bin3$ is 500Mbit/s.

To evaluate the qualities of these random numbers, we first
calculated the autocorrelation of each random train. The results are
shown in Fig.5. From Fig.5, we can see that the residual correlation
of $Bin3$ is significantly lower than that of $Bin1$.

\begin{figure}[!t]\center
\resizebox{9.5cm}{!}{\includegraphics{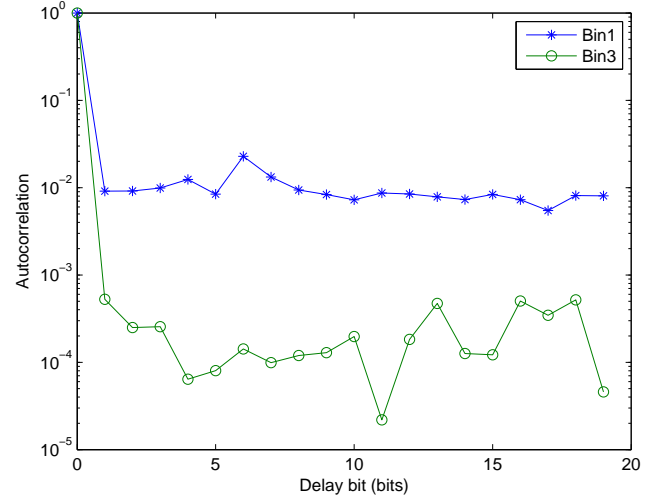}}
\caption{Autocorrelations of the random number trains acquired at
1Gbit/s. Note for Bin3, the equivalent random number generation rate
is 500Mbit/s.}
\end{figure}

We further test the randomness of $Bin3$ with the DIEHARD test suite
\cite{DIEHARD}. As shown in Table 1, $Bin3$ passed all the tests.

\begin{figure}[!t]\center
\resizebox{8.5cm}{!}{\includegraphics{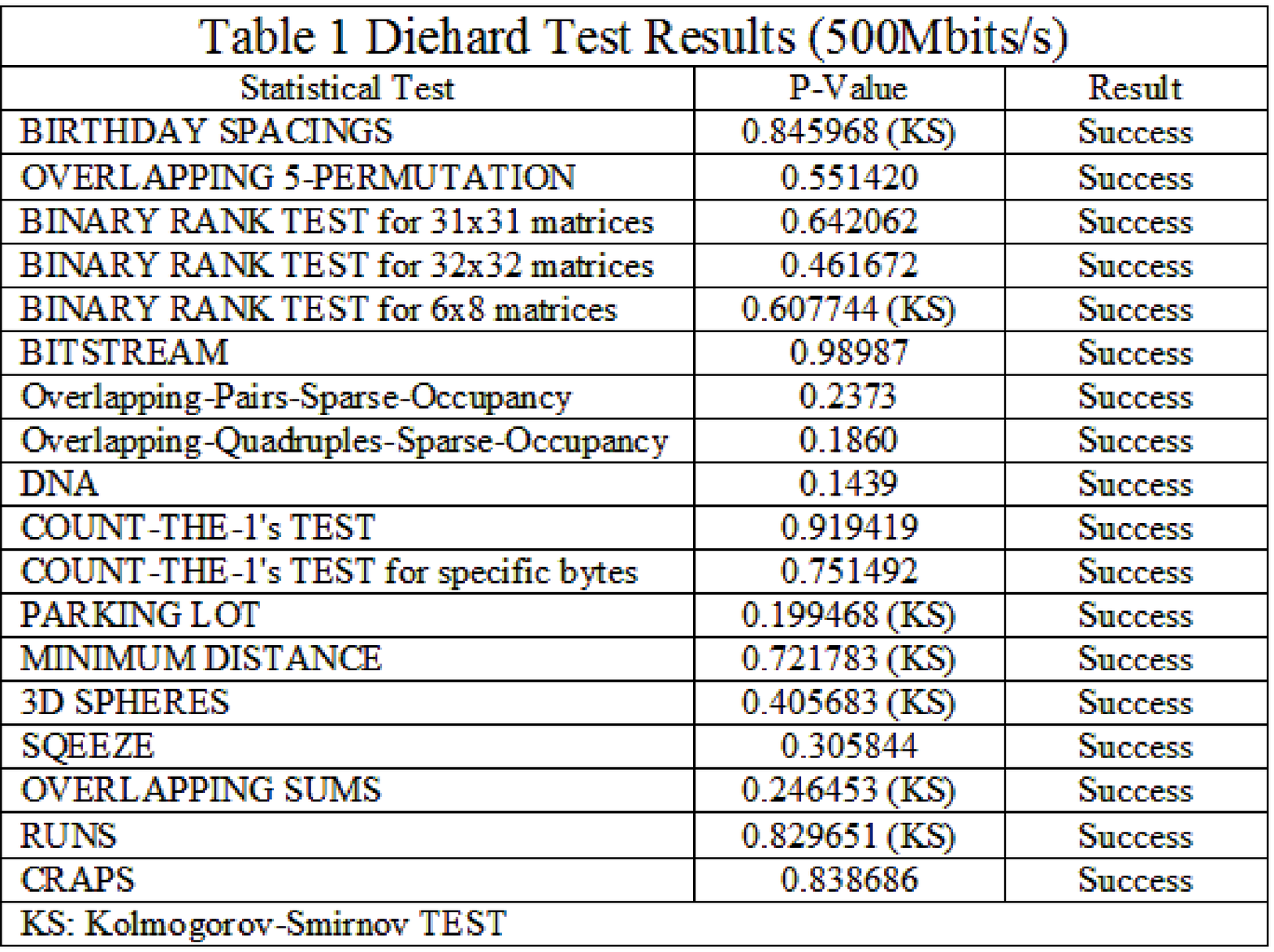}}
\end{figure}

\section{Discussion and Conclusion}

In this paper, we have presented a high speed random number
generation scheme based on measuring the quantum phase noise of a
DFB laser diode. The whole system is constructed with off-the-shelf
components and a random number generation rate of 500Mbit/s has been
demonstrated. Currently, the random number generation rate is mainly
limited by the speeds of the oscilloscope (3GHz bandwidth) and the
photo detector (5GHz bandwidth). By employing a detection/sampling
system with a larger bandwidth and a higher sensitivity, we believe
higher random number generation rate is achievable.

Comparing with the random number generation scheme based on
measuring vacuum noise \cite{Trifonov07}, the scheme we proposed
here is realized by interfering two relatively strong laser beams.
The photo-detector does not have to be shot noise limited. So it is
easier to implement.

There are also similarities and differences between our scheme and
the one reported in \cite{Uchida08}. The SE contributes to both
amplitude fluctuation and phase fluctuation. However, for a laser
operated under normal conditions, it is very difficult to resolve
the small amplitude fluctuation due to SE. In \cite{Uchida08}, the
authors operated the lasers under chaotic conditions by introducing
strong external feedbacks. The observed noise is mainly due to the
chaotic behavior of the lasers rather than the quantum noise from
SE. In contrast, the quantum phase fluctuation can readily be
measured with a conventional interferometric setup, as we have shown
in this paper.

There is a lot of room to further improve the QRNG presented here.
The sensitivity of the detection system can be further improved by
replacing the photo detector with a balanced detector followed by an
electrical substraction circuit; the real time oscilloscope can be
replaced by either a high speed comparator or a high speed analogue
to digital convertor; the DFB laser used in the current system could
be replaced by a combination of a broadband light source and a
narrowband optical filter. In this case, the coherence time (or
linewidth) is determined by the bandwidth of the filter.

We would like to end this paper with some general comments on RNGs
for secure communications.

Typically, a true RNG consists of two components: a high entropy
source and a randomness extractor \cite{Barak03}. The high entropy
source could be a physical device whose output is more or less
unpredictable, while the randomness extractor could be an algorithm
which generates nearly perfect random numbers from the output of the
high entropy source. For example, in the QRNG described in this
paper, the combination of the laser source and the delayed
self-heterodyning system can be treated as a high entropy source,
while the bitwise XOR operation, which has been adopted to improve
the randomness, can be treated as a randomness extractor. Note the
randomness extractor normally generates a short, nearly perfect
random number train from a long, imperfect one. Obviously, to design
an appropriate randomness extractor, we need to know the entropy of
the source in advance.

The entropy of a practical device originates from both quantum noise
and classical noise. As we have discussed in Section I, true random
numbers with proven randomness can only be generated from
irreducible quantum randomness. Thus it is important to quantify
with respect to the observed entropy, how much is contributed by the
quantum noise. In \cite{Fiorentino07}, the authors proposed a scheme
to quantify the ``min-entropy'' (or the irreducible quantum entropy)
of a two dimensional quantum system by employing quantum state
tomography. This is an interesting idea. However, it does not take
into account the classical noise contributed by the detection
system. Moreover, it is not clear how to apply this idea to
continuous variables.

Another interesting topic is how to deal with the finite response
time $T_R$ of a practical detection system. Normally, we assume that
the resulting correlation between adjacent samples drops
exponentially when we increase the sampling period $T_S$. In
practice, this correlation can be neglected by using a $T_S$ much
larger than $T_R$. However, this will reduce the random number
generation rate. To achieve the highest random number generation
rate, it might be more efficient to tolerate a finite correlation at
the sampling stage and let the randomness extractor to remove the
residual correlation later on.

Finally, in the special case of QKD, the randomness extractor might
be integrated into the privacy amplification process. In QKD, after
the quantum transmission stage, the two users need to perform error
correction and privacy amplification (to correct errors and remove
the eavesdropper's information) on the raw key to generate the final
secure key. Since the data size of the raw key is much less than
that of the random numbers used in the QKD experiment, it might be
much efficient to treat the imperfection of the QRNG as partial
information leaked to the eavesdropper, which can be removed during
privacy amplification.

We are thankful for enlightening discussions with Fei Ye and Amr S.
Helmy. Support of the funding agencies CFI, CIPI, the CRC program,
CIFAR, MITACS, NSERC, OIT, and QuantumWorks is gratefully
acknowledged.

Notes Added: after the completion of a preliminary version of this
paper, we notice that a preprint \cite{Hong09} has recently been
posted on quant-ph.


\begin{thebibliography}{00}
\bibitem{bell}
The testing of Bell inequalities also requires random numbers.
Hidden correlations between the supposedly local random numbers can
lead to a loophole in the testing of Bell inequalities. In fact,
people have proposed generating the random numbers for testing Bell
inequalities by observing lights from distant stars in different
sectors of the sky. However, the inflation theory shows that even
different sectors of the sky might previously been in causal
contacts with each other and, therefore, their star lights might
still have hidden correlations.
\end{thebibliography}
\end{document}